# Copy-Paste Image Augmentation with Poisson Image Editing for Ultrasound Instance Segmentation Learning

Wei-Hsiang Shen
*Dept. of Electrical Engineering*
*National Tsing Hua University*
Hsinchu, Taiwan
whshen@gapp.nthu.edu.tw

Meng-Lin Li
*Dept. of Electrical Engineering*
*Inst. Of Photonic Technologies*
*Brain Research Center*
*National Tsing Hua University*
Hsinchu, Taiwan
mlli@ee.nthu.edu.tw

*Abstract*—Deep learning has shown great success in high-level image analysis problems; yet its efficacy relies on the quality and diversity of the training data. In this work, we introduce a copy-paste image augmentation for ultrasound images. The Poisson image editing technique is used to generate realistic and seamless boundary transitions around the pasted image. Results showed that the proposed image augmentation technique improves training performance in terms of higher objective metrics and more stable training results.

*Keywords—deep learning, image augmentation, Poisson image editing, ultrasound image.*

## I. Introduction

Deep learning techniques have greatly succeeded in high-level image analysis problems, enabling better computer-based visual understanding. Within the domain of ultrasound images, deep learning has been employed for many tasks, including image classification, reconstruction, segmentation, and ultrasound beamforming [1]. One pivotal factor determining the effectiveness of a deep learning model is the quality and diversity of its training data. However, for ultrasound images, clinical data is challenging to collect, and its annotation requires expert knowledge which is time-consuming. In addition, the collected data may often lack the diversity necessary to include various lesions, limiting the model's ability to generalize across diverse clinical cases.

A straightforward approach to increase the amount of data is to use image augmentation. This technique involves applying random modifications to the original annotated images to create additional image pairs. Such modifications typically include cropping, flipping, geometric transformation, and brightness adjustment. However, although conventional image augmentation generates a larger dataset, it does not introduce data with higher diversity.

Recent studies have proposed advanced image augmentation techniques to generate data with increased diversity, such as mixup, cutout, cutmix, and copy-paste augmentation [2]-[5]. Others have extended the concept to different medical image modalities using the domain knowledge of those modalities, such as Tensor Mixup for MRI images and Stain Mixup for histopathology images [6], [7].

In this work, we propose a novel copy-paste image augmentation approach for ultrasound images that utilizes the Poisson image editing technique to produce realistic and diverse training samples. Unlike conventional copy-paste methods, which often introduce artificial boundaries in ultrasound images and thus hinder training performance, we use Poisson image editing for image pasting to create smooth boundary transitions [8]. Our results show that the proposed technique could not only create realistic samples but more importantly, improve the training performance.

## II. Materials and Methods

To increase the number of training samples, we use copy-paste augmentation technique. Figure 1 shows the detailed procedure of our proposed method. A source region of interest (ROI) is randomly chosen and cropped based on its annotation mask. Then, several random geometric transformations are applied onto the source region of interest (ROI), such as scaling, flipping, and rotation. Finally, the cropped ROI is pasted onto another ultrasound image at a random location with Poisson image editing.

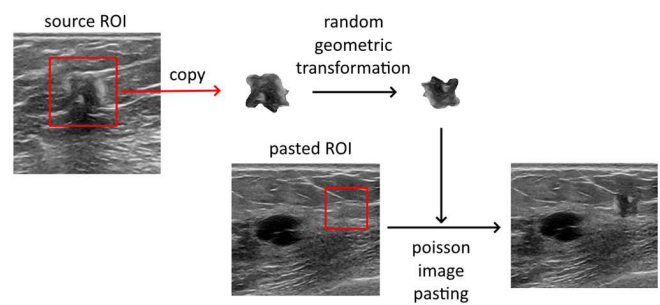

Fig. 1. The pipeline of the proposed technique. A source ROI is chosen and cropped based on its annotation mask, and then random geometric transformation is applied onto the ROI. Finally, the cropped ROI is pasted onto another backgraound image with Poisson image editting technique.

Simply pasting the ROI directly can introduce artificial boundaries, leading to unrealistic images. To address this issue, we employ Poisson image editing, a gradient-domain image processing technique. This method matches the boundary of the pasted image with the background and generates smooth gradient transition that seamlessly blends with the surrounding area, allowing us to paste each ROI in every location of the image.

*A. Poisson Image Editing*

Poisson image editing is a gradient-domain image processing technique that involves solving a Poisson equation with specified boundary conditions [8]. In this work, the pasting of images is framed as an interpolation problem, where the goal is to interpolate meaningful content within the pasted ROI. The interpolation problem can be described as an optimization problem,

$$\min_f \iint_\Omega |\nabla f|^2 \, d\Omega \text{ with } f|_{\partial\Omega} = f^*|_{\partial\Omega} \quad (1)$$

where $\Omega$ is the region of pasted ROI, $d\Omega$ is the boundary of region $\Omega$, $f$ is the unknown interpolant in region $\Omega$, $f^*$ is the known background image, and $\nabla$ is the gradient operator. This is the simplest form of interpolation, where the gradient of the interpolant is minimized while maintaining fixed image boundaries as the boundary condition. However, this simple approach would result in blurry interpolant without meaningful content.

Therefore, a guidance field $\boldsymbol{v}$ is introduced to guide the interpolant toward certain content.

$$\min_f \iint_\Omega |\nabla f - \boldsymbol{v}|^2 \, d\Omega \text{ with } f|_{\partial\Omega} = f^*|_{\partial\Omega} \quad (2)$$

In our case, the guidance field is the gradient of the source ROI to effectively guide the interpolant to the source ROI. The minimization problem can be solved by the unique solution following the Poisson equation with Dirichlet boundary conditions.

## III. RESULTS

To assess the effectiveness of our proposed image augmentation method, we utilized the Breast Ultrasound Dataset B [9]. Figure 2 shows a breast ultrasound image pasted with cysts using the proposed technique. Cyst (a) is an actual cyst in the original image, whereas cyst (b) and (c) are manually pasted. The boundary transition is smooth and seamless, and the cysts look realistic. In addition, the proposed technique could paste ROI on low-intensity location (c) without artifacts.

We trained a U-Net on the dataset for instance segmentation to predict the masks for each cyst [10]. K-fold validation (K=5) is used to evaluate the performance of the trained networks. The F1 score is used as the objective metric to evaluate the performance of instance segmentation. The results show improved performance and more stable training results. Specifically, compared to only using conventional augmentation techniques, the proposed technique yielded an average F1 score increased from 0.48 to 0.67, as shown in Figure 3.

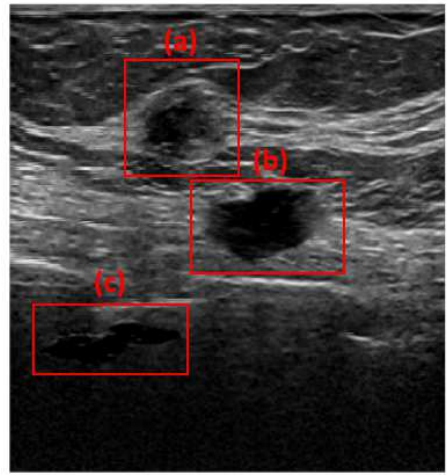

Fig. 2. An example result of the proposed image augmentation technique. (a) is an actual cyst, whereas (b) and (c) are manually pasted through Poisson image editing. The boundary transition is seamless, making the pasting realistic.

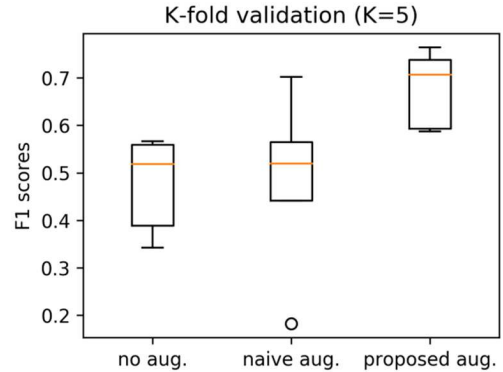

Fig. 3. K-fold validation results of training with no augmentation, conventional augmentation, and the proposed augmentation technique. The proposed technique yields higher performance in terms of F1 score and more stable training results.

## IV. CONCLUSIONS

In this study, we present a novel copy-paste image augmentation technique for ultrasound images using the Poisson image editing technique. The proposed technique can generate realistic images and seamless boundary transitions around the pasted image. Results showed improved F1 scores and more stable training results.

## ACKNOWLEDGMENTS

This research is supported by National Science and Technology Council, Taiwan (MOST 110-2221-E-007-011-MY3)

## REFERENCES

[1] Y. Wang *et al.*, "Deep learning in medical ultrasound image analysis: a review." *IEEE Access*, vol. 9, pp. 54310– 54324, 2021.
[2] H. Zhang, M. Cisse, Y. N. Dauphin, and D. Lopez-Paz, "mixup: Beyond empirical risk minimization." *arXiv*:1710.09412, 2017.

[3] S. Yun *et al.*, "Cutmix: Regularization strategy to train strong classifiers with localizable features." *Proceedings of the IEEE/CVF international conference on computer vision.* 2019, pp. 6022–6031.

[4] T. DeVries and G. W. Taylor. "Improved regularization of convolutional neural networks with cutout." *arXiv*:1708.04552, 2017.

[5] G. Ghiasi *et al.*, "Simple copy-paste is a strong data augmentation method for instance segmentation." *Proceedings of the IEEE/CVF conference on computer vision and pattern recognition.* 2021, pp. 2918–2928.

[6] Y. Wang, and Y. Ji, "TensorMixup Data Augmentation Method for Fully Automatic Brain Tumor Segmentation." *26th International Conference on Pattern Recognition (ICPR)* 2022, pp. 4615–4622..

[7] J.-R. Chang *et al.*, "Stain mix-up: Unsupervised domain generalization for histopathology images." *Medical Image Computing and Computer Assisted Intervention–MICCAI* 2021, pp. 117–126.

[8] P. Pérez, M. Gangnet, and A. Blake. "Poisson image editing." *ACM SIGGRAPH* 2003, pp. 313-318.

[9] M. H. Yap et al., "Automated breast ultrasound lesions detection using convolutional neural networks." *IEEE journal of biomedical and health informatics*, vol. 22, pp. 1218-1226, 2017.

[10] O. Ronneberger, P. Fischer, and T. Brox. "U-net: Convolutional networks for biomedical image segmentation." Medical Image Computing and Computer-Assisted Intervention (MICCAI), vol. 9351, pp. 234–241, 2015.